\documentclass[reprint,superscriptaddress,showpacs,amsmath,amssymb,prx]{revtex4-1}

\usepackage{bm}
\usepackage[mathlines]{lineno}
\usepackage{mathtools}
\usepackage{natbib}
\usepackage[shortcuts]{extdash}
\usepackage{newtxtext}

\usepackage{siunitx}
\def\bP{\mathbb{P}}
\usepackage{bbm}
\usepackage{dsfont}
\usepackage{array}
\usepackage{xcolor}
\usepackage[normalem]{ulem}
\usepackage{subfigure} 
\usepackage{amsmath} 
\usepackage{amsfonts}
\usepackage{booktabs}
\usepackage{tabularx}

\usepackage{amsthm}
\usepackage{amsmath} 
\usepackage{amsfonts}
\usepackage{amssymb}
\def\bP{\mathbb{P}}
\usepackage{bbm}
\usepackage{dsfont}
\usepackage{graphicx}
\usepackage{hyperref}

\providecommand{\vect}[1]{%
  {\boldsymbol{#1}}%
}

\let\epsilon\varepsilon

\begin{document}

\title{Physically-inspired computational tools for sharp detection\\ of material inhomogeneities in magnetic imaging}

\author{Illia Horenko}
\affiliation{Universit\`a della Svizzera Italiana, Faculty of Informatics, Via G. Buffi 13, TI-6900 Lugano, Switzerland}

\author{Davi Rodrigues}
\affiliation{Johannes-Gutenberg University Mainz, Institute of Physics, Staudinger Weg 9, 55128 Mainz, Germany}

\author{Terence O'Kane}
\affiliation{Climate Forecasting, CSIRO Oceans and Atmosphere, Castray Esplanade, 7001 Hobart, Tasmania}

\author{Karin Everschor-Sitte}
\affiliation{Johannes-Gutenberg University Mainz, Institute of Physics, Staudinger Weg 9, 55128 Mainz, Germany}

\date{\today}

\begin{abstract}
Detection of material inhomogeneities is an important task in magnetic imaging and plays a significant role in understanding physical processes. For example, in spintronics, the sample heterogeneity determines the onset of current-driven magnetization motion. While often a significant effort is made in enhancing the resolution of an experimental technique to obtain a deeper insight into the physical properties, here we want to emphasize that an advantageous data analysis has the potential to provide a lot more insight into given data set, in particular when being close to the resolution limit where the noise becomes at least of the same order as the signal.
In this work we introduce two tools - the average latent dimension and average latent entropy - which allow for the detection of very subtle material inhomogeneity patterns in the data. For example, for the Ising model, we show that these tools are able to resolve exchange differences down to 1\%. For a micromagnetic model, we demonstrate that the latent entropy can be used to detect changes in the easy axis anisotropy from magnetization data.
We show that the latent entropy remains robust when imposing noise on the data, changing less than $0.3\%$ after adding Gaussian noise of the same amplitude as the signal. 
Furthermore, we demonstrate that these data-driven tools can be used to visualize inhomogeneities based on MOKE data of magnetic whirls and thereby can help to explicitly resolve impurities and pinning centers. To evaluate the performance of the average latent dimension and entropy, we show that they 
outperform common instruments ranging from standard statistics measures to state-of-the art data analysis techniques such as Gaussian mixture models not only in recognition quality but also in the required computational cost.
\end{abstract}

\pacs{ }

\maketitle


\section{Introduction}

In the application-driven field of spintronics sample quality control is of a particular importance for the effectiveness of a device~\cite{Wolf2001,Sato2002,Wu2012,Dietl2014,Linder2015}.
For example, by engineering the layering structure of Magnetic Tunnel Junctions (MTJ's) it was possible to enormously increase the tunneling magneto resistance (TMR) ratio from originally a few percent \cite{Julliere1975} to sizable ratios which can be explored for devices~\cite{Grunberg1986, Baibich1988, Parkin2004}. This allowed for a revolution in computer industry as MTJs became the core of the magnetic random access memory (MRAM) technology~\cite{Tehrani2000,Nishimura2002,Gallagher2006}.
Also when moving magnetic textures such as domain walls and skyrmions the quality of the sample and its external conditions, such as temperature gradients, play an important role in their dynamics~\cite{Metaxas2007,Schulz2012,Kong-2013aa,Hanneken2016,Litzius2020}. For example, skyrmions as magnetic particle-like topological whirls, have been shown to be able to elegantly move around obstacles and impurities~\cite{Iwasaki2013a,Sampaio2013}. As such, being able to fully predict the motion / trajectories of skyrmions in a device, it is important to know the sample properties and detect potential inhomogeneities which are unavoidable in any real sample. Beyond single trajectory dynamics the effect of impurities is also crucial on their thermodynamic properties~\cite{Schulz2012}. While - theoretically - in a clean and thus effectively translationally-invariant sample the length of a mean free path of a magnetic skyrmion should grow linearly with increasing temperature \cite{Schutte2014a}, experiments show the length of a mean free path of a magnetic skyrmion does grow exponentially due to the presence of impurities~\cite{Zazvorka2019}. 
Furthermore, impurities can also be the nucleation center of magnetic structures~\cite{Sampaio2013,Sitte2016,Everschor-Sitte2016,Hanneken2016,Stier2017,Buttner2017a}.
Thus, having a thorough understanding of impurities and inhomogeneities might not only allow to better predict the physics but also helps engineering better devices. 

In this article we challenge the data-driven machine learning approaches with a task of resolving very tiny inhomogeneities in magnetization data. We present results for systems of increasing complexity, from controlled model system simulations towards experimental data. We introduce two scalable inference tools - the \emph{latent entropy} and the \emph{latent dimension} -   in Sec.~\ref{sec:theory} of this manuscript. In Sec.~\ref{ssec:Ising}, we start by introducing the 2D inhomogeneous Ising model and show that using these tools it is possible to disentangle exchange differences of down to 1\%. In Sec.~\ref{ssec:MicroMag} we present results for a micromagnetic Landau-Lifshitz-Ginzburg model and finally, in Sec.~\ref{ssec:ExpMag}, for noisy experimental data.
Comparing various common data analysis tools we show that the quality of latent feature detection should be  considered together with a computational scalability of the method.
In comparison to common data analysis methods, the latent entropy and latent dimension provide a cheap and robust possibility for detecting inhomogeneities from very large data sets and to extract patterns that are hidden in noisy  time-resolved measurements.
It is demonstrated that the latent tools (described in the Sec.~~\ref{sec:theory}) improve data analysis for the magnetic model systems and outperform the most common measures such as the mean, the variance, the autocorrelation, the Gaussian mixture models (GMMs) for the considered magnetic imagining applications. We provide a mathematical proof that the computational iteration costs and memory requirements of the introduced tools are independent of the data statistics size and the original data dimension. 
Furthermore, we quantify and make visible the bias imposed on experimental data by common video-compression methods which are frequently used to document the MOKE experiments.

\begin{figure*}
\begin{center}
    \includegraphics[width=1.0\textwidth]{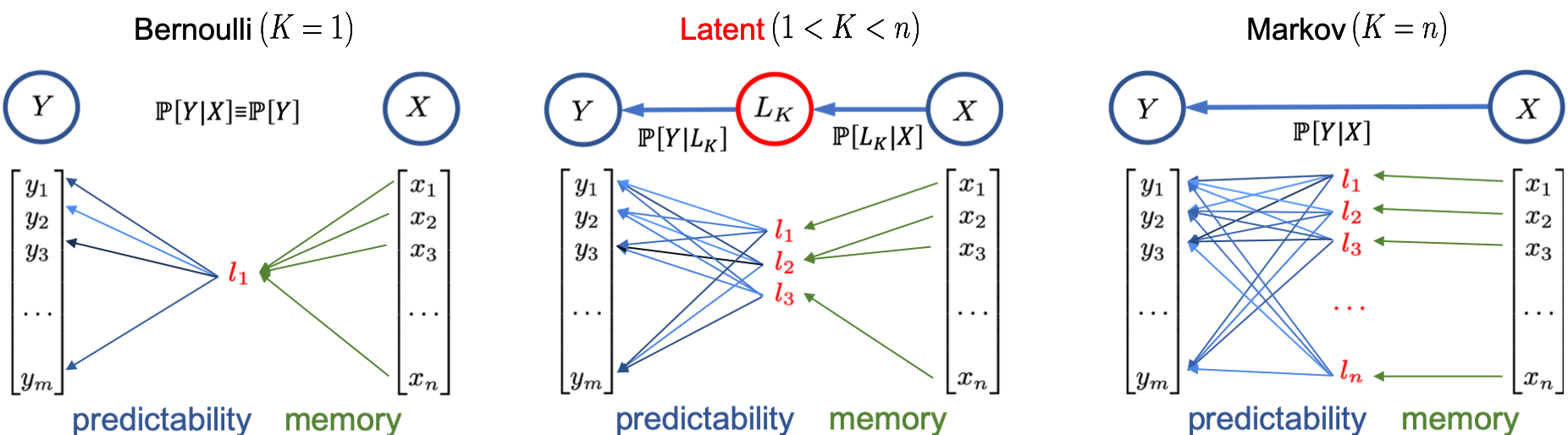} 
    \caption{ {\bf Possible relations between discrete data sets $X$ and $Y$} through the potentially hidden variable $L_K$ with latent dimension $K$. For a Bernoulli (Markov) process the latent dimension is $K=1$ ($K=n$). On the lower pannels we have schematics of the transition probabilities for different values of $K$. Different shades for the transition probabilities represent different probabilities. Small latent dimension $K$ leads to less memory of your system, i.e. $Y$ is more independent of $X$.}
    \label{fig:algorithm}
\end{center}
\end{figure*}  

\section{Latent entropy and latent dimension measures for detection of patterns in time series data}
\label{sec:theory}

In the following we introduce two physically motivated computational inference tools -- the latent entropy and latent dimension measures. These two cost-effective tools can be applied to pattern recognition in video data. \footnote{While in this work we focus on recognizing inhomogeneities in magnetic samples, we provide the analysis of a non-magnetic example on a completely different length-scale in the App.~\ref{app:andromeda}, where we examine an amateur movie of the Andromeda galaxy.} 
The underlying idea of the new tools is that we consider the time evolution of each spatial data point in time individually. Typically, a data point in a movie data series corresponds to one (or a patch of) pixel of that movie, which can assume a certain values over time. 
 By performing a statistical analysis of the transition probability between the discretized values of each spatial data point in two consecutive movie frames over time, we infer information from the underlying physics at each data point. The properties and degrees of freedom of the resulting transition matrix are related to the number of possible latent (hidden) processes and reveal the constraints and expected behaviours of the observed system.
The correlation between the initial and consecutive state gives information about the memory of the system while the tendency of system to remain within a certain range encodes its stochasticity. For example, for a Bernoulli process, consecutive states are completely independent, while for a Markov process the initial state is important.
More generally, the transition between two consecutive states $X$ and $Y$ can be related via unobserved latent processes $L_K$, see Fig.~\ref{fig:algorithm}.
Based on this picture we propose two tools - the latent dimension and latent entropy - which reveal information about the memory and stochasticity, respectively. 
Relying on thermodynamic input, we show below that the introduced tools are cost-efficient and extremely sensitive to the underlying physics  - while remaining robust with respect to the noise, even when the noise amplitudes imposed on the original signal become as large as the amplitudes of the signal.

In subsection ~\ref{subsec:Setup} we describe how to calculate the latent dimension and latent entropy for each data point. And in subsection \ref{subsec:Properties} we study the properties of these two measures.

\subsection{Setup and algorithm to compute latent relation measures}
\label{subsec:Setup}

The algorithm to compute the latent measures is a three step process which is performed for each data point of the movie. To explain the algorithm we will use a following notation. 
At a certain time $t$ of the movie, each data point assumes one of the $n$ discrete values $x=\{x_1,x_2,\dots,x_n\}$ \footnote{For common ways of discretization and obtaining the optimal set of categories from (continuous) data we refer to Refs.~\cite{Macqueen1967, Hartigan1979, Manning2008, kurgan04,kmeans10}.} and we label the corresponding data sequence as $X=\{X(t=1),X(t=2),\dots,X(t=N)\}$, where $(N+1)$ is a total number of time frames in the movie.
For the consecutive movie frames we assign the set $Y=\{Y(t=1),Y(t=2),\dots,Y(t=N)\}$ with $Y(t) = X(t+1)$.

For a certain data point at a time t,  transition from a value $x_i$ via a latent process to a value $y_j$ is described by the conditional probabilities and the \emph{exact law of the total probability} \cite{Gardiner2004} 
\begin{multline}
\label{eq:mast_reduced}
\bP[Y(t) = y_{j}] = \sum_{k=1}^{K} \bP[Y(t) = y_{j}, L_{K} = l_{k}]\times\\ \times \bP[L_{K} = l_{k}, X(t) = x_{i}] \, \bP[X(t) = x_{i}],
\end{multline}
as depicted in Fig.~\ref{fig:algorithm}. Here $\bP[A=a_i]$ describes the probability of variable $A$ to assume value $a_i$ and $\bP[A=a_i, B=b_j]$ denotes the conditional probability 
of $A$ assuming $a_i$ while $B$ is assuming $b_j$.
The unobserved latent process $L_{K}(t)$ takes values from the $K$ latent categories $\textit{l}=\{\textit{l}_{1},\textit{l}_2,\dots,\textit{l}_K\}$ \cite{hofmann99,hofmann01}. 
A number of efficient algorithms to infere the matrix 
\begin{equation}\label{eq:LambdaK}
(\Lambda_K)_{ji} \equiv \sum_{k=1}^{K} \bP[Y = y_{j}, L_{K} = l_{k}] \bP[L_{K} = l_{k}, X = x_{i}]
\end{equation}
for given data sequences $X$ and $Y$  was developed in context of the Probabilistic Latent Semantic Analysis models (PLSA) and the Direct Bayesian Model Reduction (DBMR) \cite{hofmann99,hofmann01,Ding06,gerber_pnas_17,gerber_scirep_18}.  These algorithms find the optimal values of $\Lambda_K$ by maximisation of the logarithm $LogL_K$ of the likelihood for the model. 
Based on these definitions the algorithm to compute the latent entropy and latent dimension follows as 
 \begin{itemize}
\item {\bf Step 1:} Compute the relation matrices $\Lambda_K$ for the given data sets $X$ and $Y$, as well as the quantities $S_K=-\frac{1}{N}LogL_K$ for every $K$ going from $1$ to $n$, see App.~\ref{app:SK}.
\item {\bf Step 2:} Compute the posterior probabilities $p_K$ for the different latent dimensions $K=1,\dots,n$ by means of the Akaike Information Criterion \cite{hurvich89} as
\begin{equation}\label{eq:AIC}
p_K=\frac{\exp\left(-\left(AICc_K-\min_K{AICc_K}\right)\right)}{\sum_{K=1}^{n}\exp\left(-\left(AICc_K-\min_K{AICc_K}\right)\right)},
\end{equation}
where $AICc_K= N S_K +V_K+\frac{V_K(V_K+1)}{N-V_K-1}$ and $V_K=\dim(\lambda_K) - K + \dim(\gamma_K)-n= (m-1)K+n(K-1)$.
\item \textbf{Step 3:} Compute the average latent entropy and the average latent dimension between $X$ and $Y$ as   \begin{equation}
\label{eq:exp_SK}
\bar{S}=\sum_{K=1}^{n}p_K S_K, \qquad \text{and} \qquad \bar{K}=\sum_{K=1}^{n}p_K K,
  \end{equation}
  and the relative latent measures as
  \begin{equation}
\label{eq:exp_rel}
\bar{S}_{rel}=\frac{\bar{S}}{S_1},
 \qquad \text{and} \qquad\bar{K}_{rel}=\frac{\bar{K}-1}{(n-1)}.
  \end{equation}
  \end{itemize}
A detailed version of the algorithm can be found in App.~\ref{app:theorydiscretel}.

To summarize, instead of specifying the latent dimension, we perform the calculation for all latent dimensions and assign to each of them the  corresponding probability of occurrence. This procedure can be viewed in analogy to the path integral formalism where all paths are allowed and weighted by their action \cite{Feynman1948}.

\subsection{Properties of the latent entropy and latent dimension}
\label{subsec:Properties}
In the following we summarize the properties of the latent measures. The corresponding mathematical theorems and their proofs can be found in a more generalized form in App.~\ref{app:theorydiscretel}.

The relative latent entropy $\bar{S}_{rel}$ and relative latent dimension $\bar{K}_{rel}$ assume values in between zero and one. 
Furthermore, they only quantify the entropy and the memory of the latent process $L$, meaning that, for example, in contrast to the total entropy, the latent entropy is not affected by the i.i.d. (independent identically distributed) noise - for example, by the Gaussian white noise - when it is added to the original data.
$\bar{S}_{rel}$ is zero if and only if the system is completely deterministic. Furthermore, 
the system is completely independent of previous states (Bernoulli process) if $\bar{K}_{rel}=0$ and $\bar{S}_{rel}=1$.

The iteration cost of computing the latent measures $\bar{S}$, $\bar{K}$, $\bar{S}_{rel}$ and $\bar{K}_{rel}$ is independent of the statistics size $N$ and the number of the data points (pixel patches) $D$ as long as $N>n^2$. It only depends on the maximal discrete latent dimension $n$ and scale as $\mathcal{O}\left(n^4\right)$, requiring no more than $\mathcal{O}\left(n^2\right)$ of memory. 

As demonstrated in the following examples, performance of the introduced latent measures  together with their minimal computational costs and memory requirements, fairly exceed 
state-of-the-art data science methods, such as the GMMs, which for example have been used to analyse the petabytes of raw data from a plethora of telescopes to obtain the first image of the black hole \cite{bouman18,bouman19}.
The computational cost of the expected latent entropies by means of the common GMMs grow linearly with the statistics size $N$ and the data dimension $D$ - and would have an iteration cost scaling of $\mathcal{O}\left(n^2ND\right)$, requiring $\mathcal{O}\left(n(N+D)\right)$ of memory, see App.~\ref{app:comparison}

Results of the numerical comparison for the full algorithm costs are shown for the Ising model in the Fig.~\ref{fig:ising}d).


\section{Data-driven Detection of Material Parameters in Heterogeneous 2D Ising models}
\label{ssec:Ising}

The Ising model is a simple toy model consisting of coupled variables on a lattice that can adopt two discrete states typically described by $s_i=\pm 1$. While originally introduced in the field of magnetism to study phase transitions \cite{Dyson1969,Wegner1971}, it is applicable in various branches of science including spin glasses \cite{Wang1990},
lattice gases \cite{Lee1952}, 
binary alloys \cite{Bortz1974},
biological systems \cite{Shi1998,Bai2010,Vtyurina2016,Jaynes1957a,Jaynes1957b,Schneidman2006},
and social applications \cite{Stauffer2008}. 
In the following we will consider the Ising model in 2D using the typical language of magnets in which the energy of the system in the absence of external magnetic fields is given by
\begin{equation}
E = -\frac{1}{2}\sum_{<i,j>}J_{ij}s_{i}s_{j}.
\label{eq:IsingEnergy}
\end{equation}
Here the spins $s_i$ on lattice site $i$ are coupled via the exchange interaction with strength  $J_{ij}$, where the sum is over pairs of adjacent spins. 
Note that for $J_{ij}>0$, the model favours the alignment of neighbouring spins.  
For a uniform coupling $J_{ij} \equiv J$ the 2D Ising model is analytically solvable \cite{Schultz1964} and it undergoes a second order phase transition at the critical temperature $T_c =  J / (k_B \ln(1 + \sqrt{2})) \sim 1.135 \, J/k_B$, with $k_B$ being the Boltzmann constant.
For temperatures lower than $T_c$ the spins get ordered, leading to a non-zero total magnetization. Above $T_c$ the temperature fluctuations are so strong that on average there is an equal number of spins up and down leading to a vanishing net magnetization.

\begin{figure*}
\begin{center}
    \includegraphics[width=1.0\textwidth]{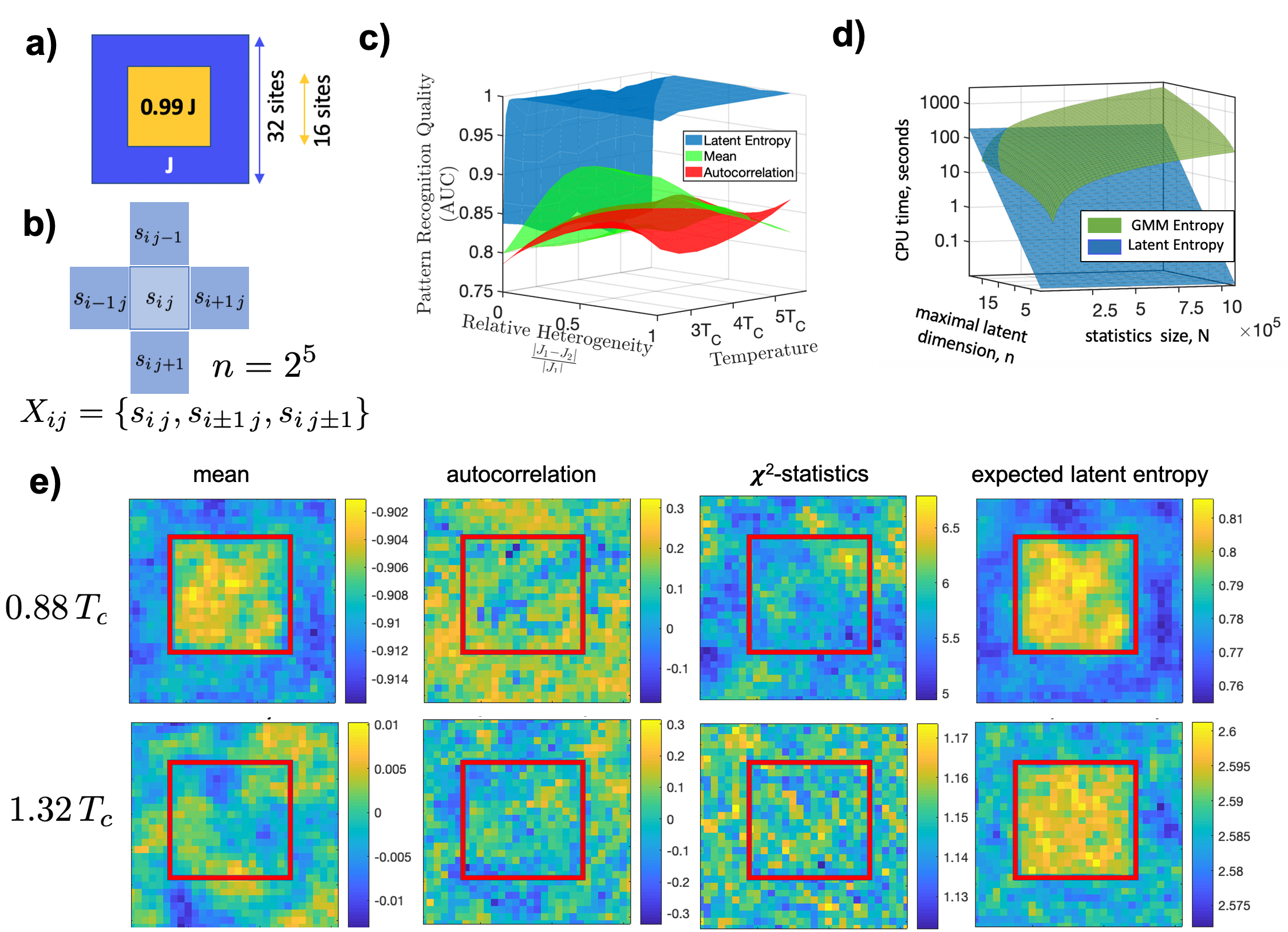}
   \end{center}
     \caption{{\bf Data analysis results for the heterogenous Ising model} 
      {\bf (a)} sketch of the corresponding model; 
      {\bf (b)} patch of $5$ lattice sites with two level states $s_{ij} =\pm1$ where $ij$ represent the position of the lattice site, yielding a total of $2^5=32$ categories;
      {\bf (c)} Area Under Curve (AUC) measure of  the heterogeneity recognition quality in the Ising model for the different data analysis tools as a function of the relative heterogeneity and temperature (in units of the critical temperature $T_c$). Values of AUC close to one indicate almost prefect recognition of the heterogeneity pattern. 
      {\bf (d)} Comparison of the mean CPU times required for the common entropy calculation  with the Gaussian Mixture Models (GMMs) and the latent entropy computation.
      {\bf (e)} Data analysis results for different methods below ($T = 0.88\, T_{c}$) and above ($T=1.32\, T_{c}$) the critical temperature. The red dotted lines indicate the inhomogeneity pattern contour. While in the ordered phase, the mean as well as the latent entropy allow for identifying the patch with a different exchange parameter, in the disordered state above $T_c$ only the latent entropy provides the desired information.}
\label{fig:ising}
\end{figure*}

We simulated a model system, shown in Fig.~\ref{fig:ising}a), where in the center the exchange coupling is reduced by $1\%$ compared to the outside value. The task of a data driven inference method is then to reconstruct the shape of an inhomogeneity region based on the simulation data. 
For the results shown in Fig.~\ref{fig:ising} we simulated $D= 32 \times 32$ lattice sites that assume values $s_{i}(t)=\pm1$ over time $t$.
We choose $J=1$ except for interactions including lattice sites in a $16 \times 16$ region in the center with $J = 0.99$. 
We consider each data point to assume $2^5=32$ different values that are obtained by considering the pixel patch shown in Fig.~\ref{fig:ising}b)

We performed $50$ randomly initialised Monte-Carlo simulations for two different regimes, corresponding to temperatures below the critical ($T= 0.88\, T_c$) and above the critical temperature ($T= 1.32\, T_c$), to obtain $N = 50\,000$ different possible configurations of the system. From the obtained  data sequences, we calculated the mean values of the magnetization, the autocorrelation, the $\chi^2$--statistics  as well as the expected latent entropy introduced in this work. 
To compute the common measures (the mean, the autocorrelation, the $\chi^2$-statistics, the mean of the square differences, the Shannon entropy from the GMM-model) we used the standard MATLAB functions (kmeans(), xcorr(), fitgmdist(), etc.). The results are shown in 
Fig.~\ref{fig:ising}e).

As expected, the common statistical tools like the mean measure work well in the regimes where either both structures order and the amplitude of the magnetization is larger in the region with enhanced coupling strength - or when one of them orders and the other one does not. This happens in the regime slightly below and around the critical temperature $T_c$. 
For temperatures above $T_c$ there is no long-range magnetic order. Spins interact and order locally within the reach of the correlation length, $\xi= \left[ (4k_B/J)|T_{c}-T|\right]^{-1}$~\cite{Pathria2011}. Therefore, the correlation length gives an upper bound for recognising magnetic order in the system and tends quickly to zero for increasing temperatures.
The correlation length for the simulations shown in Fig.~\ref{fig:ising} is on the order of one lattice spacing. As such the resulting states are disordered and have significant fluctuations. This hinders considerably the ability to recognise underlying patterns by the usual methods.

Our results show that the scalable latent measures can identify the change in parameters by analysing the latent transition probabilities, both in the ordered and disordered phase.
We evaluated the pattern recognition quality by calculating the Area Under Curve (AUC), which compares the identifiable area with the known area assigned for the system, for each of the measurements, see Fig.~\ref{fig:ising}c) and show that the latent measures by far exceeds the others in performance, while being computationally cheaper, see Fig.~\ref{fig:ising}d). AUC values close to 1.0 refer to a perfect recognition of the inhomogeneity pattern, wheres AUC close to 0.5 corresponds to the cases where the tool is completely failing to recover the pattern.
Another important remark to validate the use of latent measures based on the latent entropy measurement above critical temperatures: from a physical point of view, entropies are continuous and well-defined functions through all possible phases of a system, including a phase transitions region. For this reason, it is a reliable quantity in the vicinity of critical temperature as well as far from it.


\section{Detecting material inhomogeneity patterns from Micromagnetic Model simulations}
\label{ssec:MicroMag}

Magnetism in the nano- and micrometer range is richer and more complex than the above two level Ising model. It is a very active research topic bearing promising, application-relevant magnetization configurations, such as domain walls \cite{Slonczewski1972,Chapman1984,Yamaguchi2004,Parkin2008},  magnetic vortices \cite{Papanicolaou1991,Shinjo2000}, and skyrmions \cite{Bogdanov1994,Muhlbauer2009a,Nagaosa2013, Everschor-Sitte2018a}.
On these length scales the atomic structures can rather be ignored, the magnetization configuration of the material can be described in a coarse-grained model and is represented as a vector field with three spatial components and a constant magnitude.

\begin{figure*}
\begin{center}
    \includegraphics[width=1.0\textwidth]{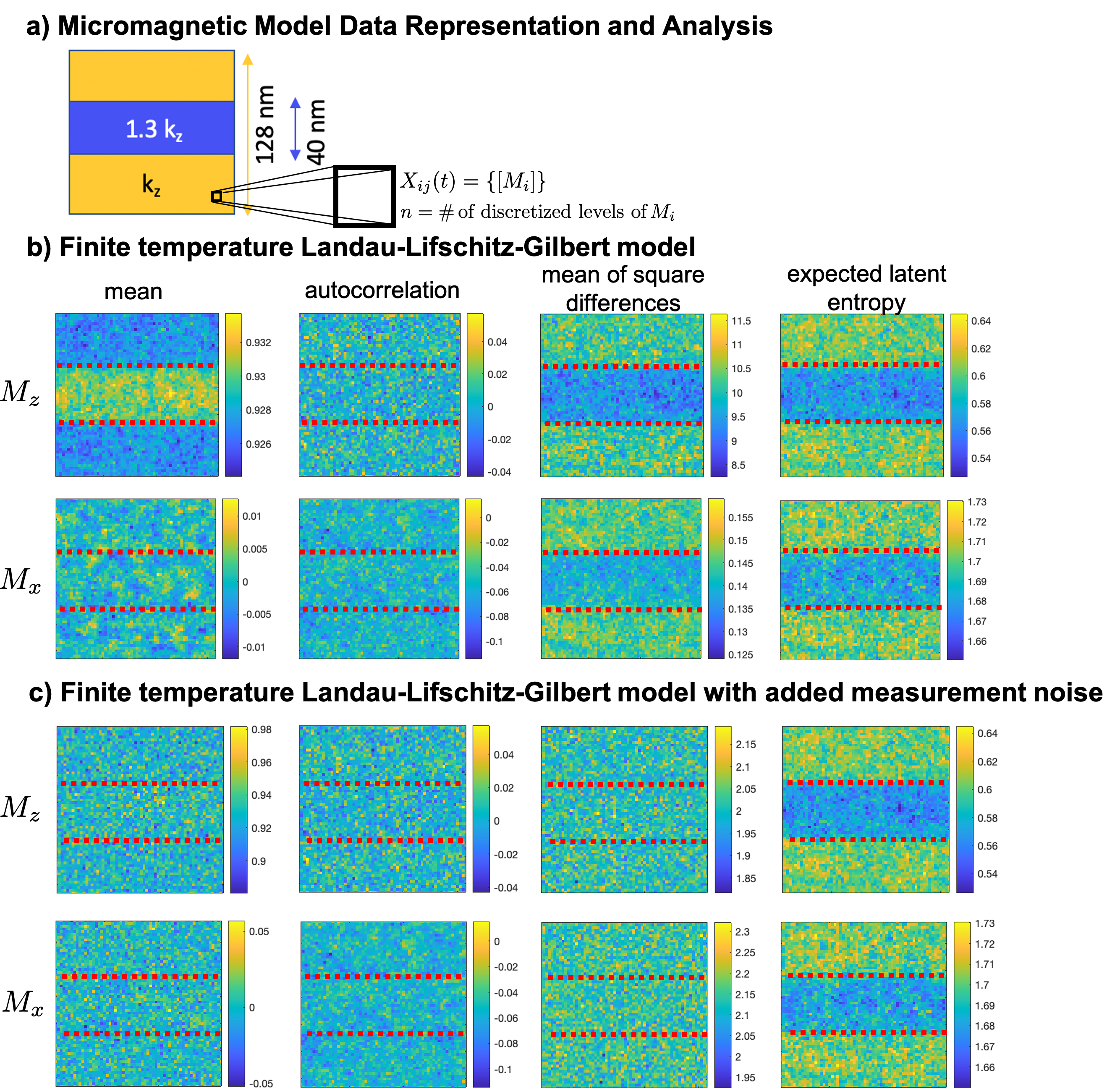}
   \end{center}
  \caption{{\bf Data analysis results for the Heterogeneous Micromagnetic Model}:  {\bf (a)} sketch of the sample configuration and discretization for pixel analysis.
  {\bf (b)} analysis of the $M_z$ (out-of-plane) and $M_x$ (in-plane) components of the magnetization in the absence of measurement noise; {\bf (c)} analysis of the $M_z$ (out-of-plane) and $M_x$ (in-plane) components of the magnetization in the presence of an added measurement noise. For all data, we calculated from the simulations results the mean value of the magnetization for each pixel over time, $\langle M_{i}\rangle$, the autocorrelation, $\langle M_{i}(t)M_{i}(t')\rangle$, the mean of the square difference, $\langle(M_{i} -  \langle M_{i} \rangle)^2\rangle$, and the expected latent entropy. The red dotted lines indicate the inhomogeneity pattern contour (a stripe in the domain middle). Notice that while in Fig.~\ref{fig:ising}, raising the temperature has increased the absolute values of the expected latent entropy, adding the measurement noise does not modify the scale of the expected latent entropy. This is because the latent entropy is directly associated to the underlying dynamics whose properties do not change with added data noise.
 }   
\label{fig:Hi-C_fit2}
\end{figure*}

We performed micromagnetic simulations using MuMax3 \cite{Vansteenkiste2014}, solving the effective equation of motion for the magnetization, i.e.\ the Landau-Lifshitz-Gilbert equation\cite{Landau1935,Gilbert2004}. 
It describes the dynamic response of the magnetization, $\vect{M}(\vect r)$, to torques and is given by 
\begin{equation}
\partial_t \vect{M} = -\gamma \vect{M}\times\vect{B}_{\textrm{eff}} + \frac{\alpha}{M_{S}} \vect{M}\times \partial_t \vect{M}.
\end{equation}
The first term on the left side is an energy conserving precessional term, where $\gamma$ is the gyromagnetic ratio, and the last term is a phenomenological damping term with strength $\alpha$. The magnitude of the magnetization is fixed to the value of the saturation magnetization, $|M(\vect r)| \equiv M_S$.
 The effective magnetic field $\vect{B}_{\textrm{eff}}$ contains the specifics of the micromagnetic model. In the presented simulations (see Fig.~\ref{fig:Hi-C_fit2}) we consider a 
$128\, \textrm{nm} \times 128\, \textrm{nm}$ square sample at room temperature, i.e.\ $T = 300\, K$,
 with $\vect{B}_{\textrm{eff}} = -2A\nabla^2 \vect{M} - 2 K M_{z} \hat{\vect{z}}  + \vect{B}_{\textrm{therm}}$ containing exchange interactions with strength $A=1.5 * 10^{-11} \,\textrm{J/m}$ and magnetic easy axis anisotropy along the out-of-plane direction with strength $K= 2.3* 10^5\, \textrm{J/m}^3$. We have increased the value of the anisotropy strength by $30\%$ in the middle region of width $40  \textrm{nm}$, see Fig.~\ref{fig:Hi-C_fit2}a). 
 The stochastic thermal field $\vect{B}_{\textrm{therm}}$ is modeled as white noise with average $\langle \vect{B}_{\textrm{therm}}(T,\vect{x},t) \rangle = 0$, and time and space correlation proportional to temperature, $\langle \vect{B}_{\textrm{therm}}(T,\vect{x},t) \vect{B}_{\textrm{therm}}(T,\vect{x}',t') \rangle \propto T \delta^{4}(\vect{x} - \vect{x}', t- t')$, where $\delta$ is the Dirac delta distribution \cite{Leliaert2017}.

The initial configuration was chosen to be the ferromagnetic ground state and then the system evolved due to the thermal fluctuations. For each $2\,\textrm{nm} \times 2\, \textrm{nm}$ cell we recorded the magnetization direction configuration $\vect M(t)$ every $50\, \textrm{fs}$. Thus, for every pixel and magnetization component we obtain a time-resolved data set (a time series), see Fig.~\ref{fig:ising}a). The values of the magnetization were discretized by means of the standard K-means discretization  technique \cite{Macqueen1967, Hartigan1979, Manning2008, kurgan04}.
Furthermore, we averaged the values over 50 randomly selected data sets from the micromagnetic model.

In Fig.~\ref{fig:Hi-C_fit2}b) we show the comparison between different statistical methods including the latent entropy $\bar{S}$ for the $z$ (out-of-plane), and $x$-component (in-plane) of the magnetization. 
We find that the latent entropy $\bar{S}$ captures the model inhomogeneity in both magnetization components.
This can be explained, as larger anisotropy leads to more ordering, and thus to higher predictability of outcomes meaning lower average latent entropy. 
We show that the latent entropy resolves the inhomogeneity region more accurately than the other statistical methods, whose accuracy may depend on the studied magnetization component.

Next, to account for the inevitably random errors in realistic measurements, we artificially added white Gaussian noise to the data. The variance of the noise was set to the same amplitude as the maximum magnetization variation. In this case, see Fig.~\ref{fig:Hi-C_fit2}c), only the latent entropy is still reliable to resolve the inhomogeneity region in both magnetization components.  Furthermore, as can be seen from the Fig.~\ref{fig:Hi-C_fit2}c), adding the noise does not change the absolute magnitudes of the expected latent entropy, as it only resolves the latent effects that are not affected by the i.i.d. noise.  Relattive difference between the latent entropy measure results obtained with and without noise in Fig.~\ref{fig:Hi-C_fit2}c) are less then 0.3\%.This feature is remarkable since we considered a noise as large as the data signal. This robustness of the latent measures can become especially useful when analysing measurements that are close to the resolution limit of a device, when the signal-to noise ratios become small.

\section{Detecting material inhomogeneity patterns from magnetization experiments}
\label{ssec:ExpMag}

\begin{figure*}[tb] 
    \includegraphics[width=1\textwidth]{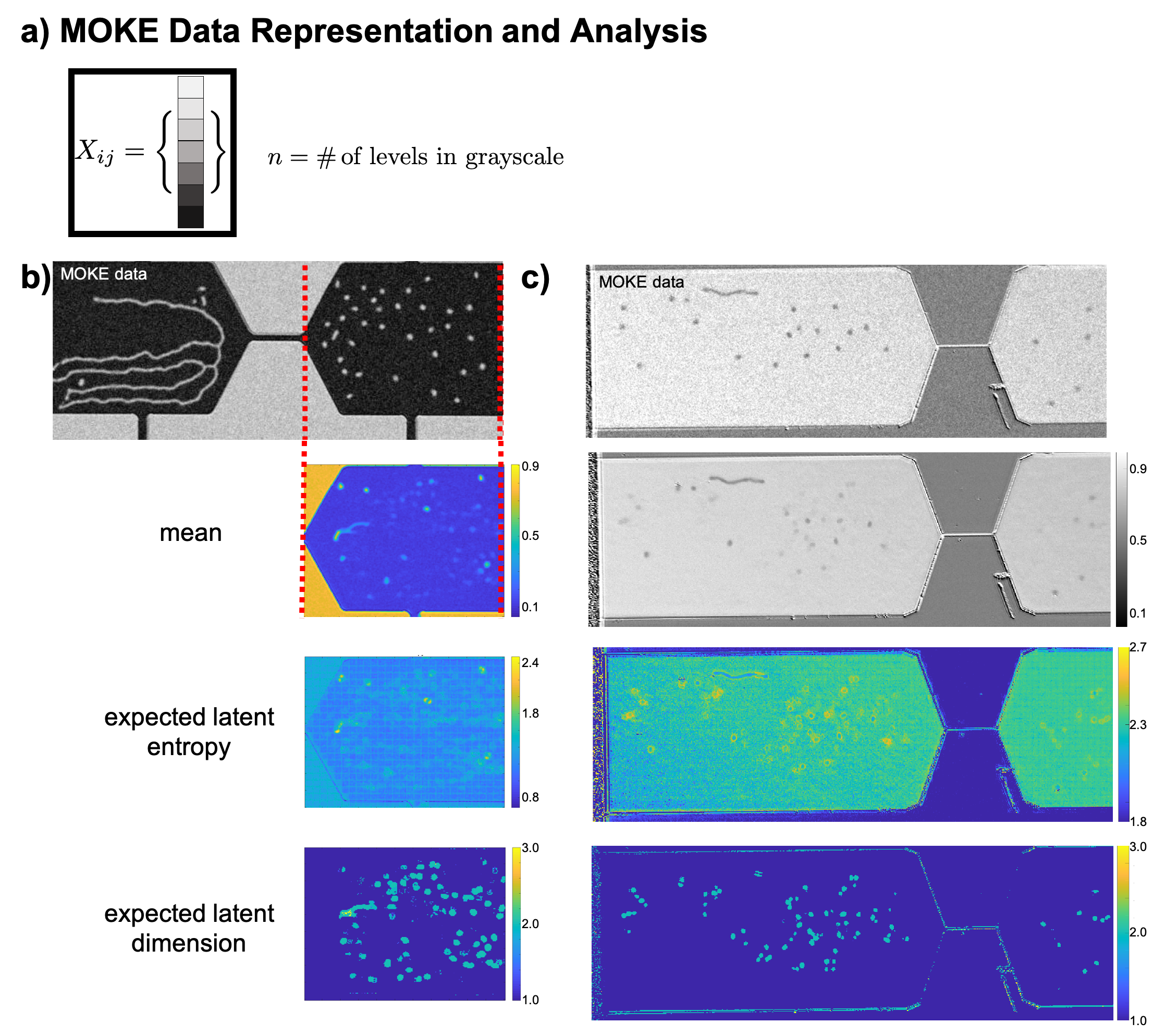} 
   \caption{ {\bf Application of data-driven tools to experimental magnetization data}: {\bf (a)} Sketch of the category (grayscale) chosen for analysing each pixel.
   {\bf (b)} Data analysis of two magnetic imaging experiments, Ref.~\cite{Jiang2015a} and Ref.~\cite{Zazvorka2019} on the left and right columns respectively. The first row shows results of magnetic structures imaged via Magneto-optical Kerr effect (MOKE) measurements. The  second, third and fourth row show the mean, the averaged latent entropy and the averaged latent dimension, respectively (see Section V for a description). The average latent entropy also reveals the effect of video compression (square lattice super-structure).
      }  
\label{fig:magnetizationdata}
\end{figure*}

Next, we analyse imaging data from magnetization dynamics experiments. The data is obtained from two different experiments\cite{Jiang2015a, Zazvorka2019} in which the out-of-plane component of the magnetization configuration was imaged with a magneto-optical Kerr effect (MOKE) microscope\cite{Hubert2009,Huang1994} and is represented at each pixel by a grey scale, as sketched in Fig.~\ref{fig:magnetizationdata}a). 
The first row in Fig.~\ref{fig:magnetizationdata}b) and c) shows one of the imaged frames in the corresponding experiment as an example. The  second, the third and the fourth row show the common statistical mean, the averaged latent entropy and the averaged latent dimension, respectively.
In both experiments, the latent entropy as well as the latent dimension show features that are not observable with the common statistical mean. In such type of magnetic systems, several factors might contribute to latent effects such as various types of material inhomogeneities, local temperature differences, etc.
Particularly,  the features accessible by the two methods are relevant to understand the dynamics of the topological magnetic whirls -- magnetic skyrmions -- such as their statistics and motion in samples.
The dynamics of topological magnetic objects are potentially interesting for the next-generation electronic devices based on spintronics principles \cite{Wolf2001}, to which the introduced methods provide relevant input.

\begin{figure} 
\centering 
    \includegraphics[width=1\columnwidth]{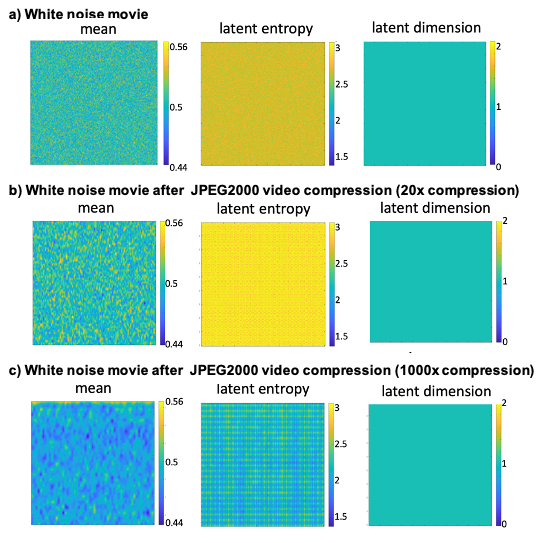} 
   \caption{ {\bf Latent entropy of the white noise movie data before and after the video compression reveals bias imposed by the common video compression tools.} Analysed movie consists of 400 image time frames, with 221x181 pixels images in every time frame. Values at every image pixel at every time frame are random realisations of the Gaussian random number generator with mean zero and variance one, obtained with the $random()$ command of MATLAB.  {\bf (a)}: Analysis of the original data (without video compression); {\bf (b)}: analysis of the mpeg-compressed movie data, with a compression factor 20; {\bf (c)}: analysis of the mpeg-compressed movie data, with a compression factor 1000.}
    \label{fig:videocompression}
\end{figure}

In the experiments analyzed in Fig.~\ref{fig:magnetizationdata}b), the authors considered a $\textrm{Ta}(5 \textrm{nm})/\textrm{Co}_{20}\textrm{Fe}_{60}\textrm{B}_{20}(\textrm{CoFeB})(1.1 \textrm{nm})/\textrm{TaO}_{x}(3 \textrm{nm})$ trilayer\cite{Jiang2015a}. With an applied electrical current, they moved worm domains - bounded regions where the magnetization is uniform and correspond to a different stable state compared to its surroundings. They moved these worm domains through a constriction and observed the creation of skyrmions at the end of it. The latent measures clearly identify inhomogeneities, where magnetic skyrmions and worm domains get strongly pinned or deflected. Detected inhomogeneity sites might explain the scattering of skyrmions and their homogeneous distribution in all directions despite the skyrmion Hall effect. Furthermore, the latent entropy also identifies a rectangular bias pattern. 
To verify the source of this rectangular grid-like pattern, we have analyzed the bias introduced on the latent entropy by common video compression tools such as MPEG. We find that an application of the latent entropy measure to a randomly generated data  acquires exactly the same bias lattice pattern when imposing the MPEG video compression  (see Fig.~\ref{fig:videocompression}). This makes sense as an MPEG compression tool basically correlates the spatial information within an effective increased pixel size, inducing changes in the predictability of the outcome. The latent dimension is, as expected, unaffected by video compression since the compression is done on individual video frames and does not affect the temporal history or memory in the data. This finding can be helpful to improve experimental data processing.

In the experiment analyzed in Fig.~\ref{fig:magnetizationdata}c), the authors studied the Brownian motion of skyrmions 
in specially tailored low-pinning multilayer material Ta(5\textrm{nm})/Co$_{20}$Fe$_{60}$B$_{20}$(1\textrm{nm})/Ta(0.08\textrm{nm})/MgO(2\textrm{nm})/Ta(5\textrm{nm}) stacks\cite{Zazvorka2019}. The analyzed data are stored as a $672 \times 510$ pixels video with the spatial resolution scale given by the measure of $305$ pixels corresponding to $50$ $\mu$m. The time step between frames is $62.5$ ms. 
This experiment aimed at studying rather homogeneous materials striving for a free motion of magnetic skyrmions 
and avoiding impurities where magnetic textures get pinned.
The time record of the skyrmions' positions, however, revealed that there are preferred positions where they tend to stay longer and which can indirectly be associated to the existence of inhomogeneities\cite{Zazvorka2019}.
As shown in Fig.~\ref{fig:magnetizationdata}c), the latent measures directly reveal these material inhomogeneities, even without having full access to all magnetization components. 
The average latent dimension sharply identifies impurities where skyrmions are more likely to be pinned. The latent entropy shows regions of distinctively-different latent entropy values, which can potentially be attributed to slightly different temperatures across the sample.
 A caveat of the data-driven measures compared in this manuscript, is that even though they are able to identify different physical patterns, they do not directly provide means to distinguish the source of the different patterns. They do, however, reveal interesting features and inhomogeneities that can then be further investigated.

The tools introduced in this work can identify changes in material parameters up to $1\%$ as shown in Sec.~\ref{ssec:Ising} for the Ising model simulations. In real experiments such as the MOKE experiments discussed in this work, the experimental data quality is, of course, also very important for the overall accuracy of the results. This includes 
the spatial and time resolution of the experimental setup as well as its sensitivity (corresponding to the resolution in different grey color values). 
The state-of-the-art of MOKE instrumentation has a spatio-time resolution in the order of $\sim1\mu$m and $\sim10^2\,$fs while they can detect rotations of the magnetization down to $\sim10\,$nrad~\cite{Henn2013,Aivazian2015,McCormick2017,Noyan2019}. We emphasize, nonetheless, that the latent measures still surpass the accuracy of other statistical tools when applied to the same data.

\section{Discussion} 
We have shown that it is possible to resolve even very subtle material inhomogeneities by means of magnetic imaging data for various magnetic systems, as the information about inhomogeneities is present in the form of latent features in the time series data. In particular we have introduced two scalable data analysis tools for the extraction of latent features, which give access to the predictability (latent entropy) and the memory functionality (latent dimension) of the system.
We have proven that the two introduced tools overcome the limitations imposed by restrictive underlying assumption of common machine learning tools (like Gaussianity and homogeneity) -  as well as the memory and computational cost scalability limitations present in popular latent inference methods like Gaussian Mixture Models. 
We have shown that these two measures outperform common tools in recognizing material inhomogeneities from magnetic imaging data. 
 For example, for the Ising model it was shown that one can resolve parameter differences of only 1\% even in the disordered phase. 
  For the micromagnetic model it was demonstrated that the time correlation is present not only in the out-of-plane but also in the in-plane component - thus providing an advantage if only one magnetization component can be measured in experiments. Moreover, in Fig.~\ref{fig:Hi-C_fit2} we have shown that the latent entropy measure is essentially not affected by Gaussian noise. Even when the noise is as large as the signal, the relative variation of the latent entropy is on the order of $0.3\%$. Thus, it is a robust measure for measuring latent features and identifying material inhomogeneities from noisy experimental data. 

We would like to compare our results also to common denoising methods exploiting spectral filtering that are typically applied to experimental data. As was proven in the fundamental work by D.~Donoho \cite{Donoho1995}, the optimal denoising can be achieved with wavelets filtering, eliminating all of the wavelet basis components of the signal whose amplitudes are below $A_c=\sigma \sqrt{2 \log(N)/N}$. Here, $\sigma$ is the variation of noise, and $N$ is the data statistics size.
As soon as $N$ is not large enough, meaning that $A_c$ becomes comparable to the amplitude of the signal, this denoising will also eliminate the underlying signal. Thus, when measuring close to the resolution limit of a device, common spectral denoising methods require an extensive amount of data for still being able to find the signal.  
For example, in Fig.~\ref{fig:Hi-C_fit2} where the signal-to-noise-ratio (SNR) is around $1.0$, common denoising would require 100 times more data to reduce the uncertainty by a factor of about $7$.
In contrast, using the latent entropy and dimension measure, that directly infer latent data structures can help reducing the amount of required data, and thus provide a path towards learning structures in small data problems.

Understanding and resolving material inhomogeneities in experimental magnetic imaging data will allow to describe samples better. Although the latent tools introduced are not able to directly identify the source of the different patterns, they provide relevant input concerning their existence and nature which can be further investigated by other methods.
As the dynamics of magnetic textures are crucially influenced by material inhomogeneities, this will provide a path towards a deeper understanding, improved prediction and engineering of the dynamics of magnetic textures.

 \section{acknowledgments}
We thank S.\ Gerber, L.\ Pospisil, J.\ Sinova, F.\ Schmid and M.\ Kl\"aui for helpful discussions.
 The work of I.H. was funded by the Mercator Fellowship in the DFG Collaborative Research Center 1114 "Scaling Cascades in Complex Systems" as well as funding from the Emergent AI Center funded by the Carl-Zeiss-Stiftung. K.E.S. and D.R. acknowledge funding from the German Research Foundation (DFG), projects EV 196/2-1, EV196/5-1 and SI1720/4-1 as well as the Emergent AI Center funded by the Carl-Zeiss-Stiftung. TJO was supported by the Australian Commonwealth Scientific and Industrial Research Organisation (CSIRO) Decadal Climate Forecasting Project (https://research.csiro.au/dfp)

\appendix

\begin{figure*} 
\centering
     \includegraphics[width=1.0\textwidth]{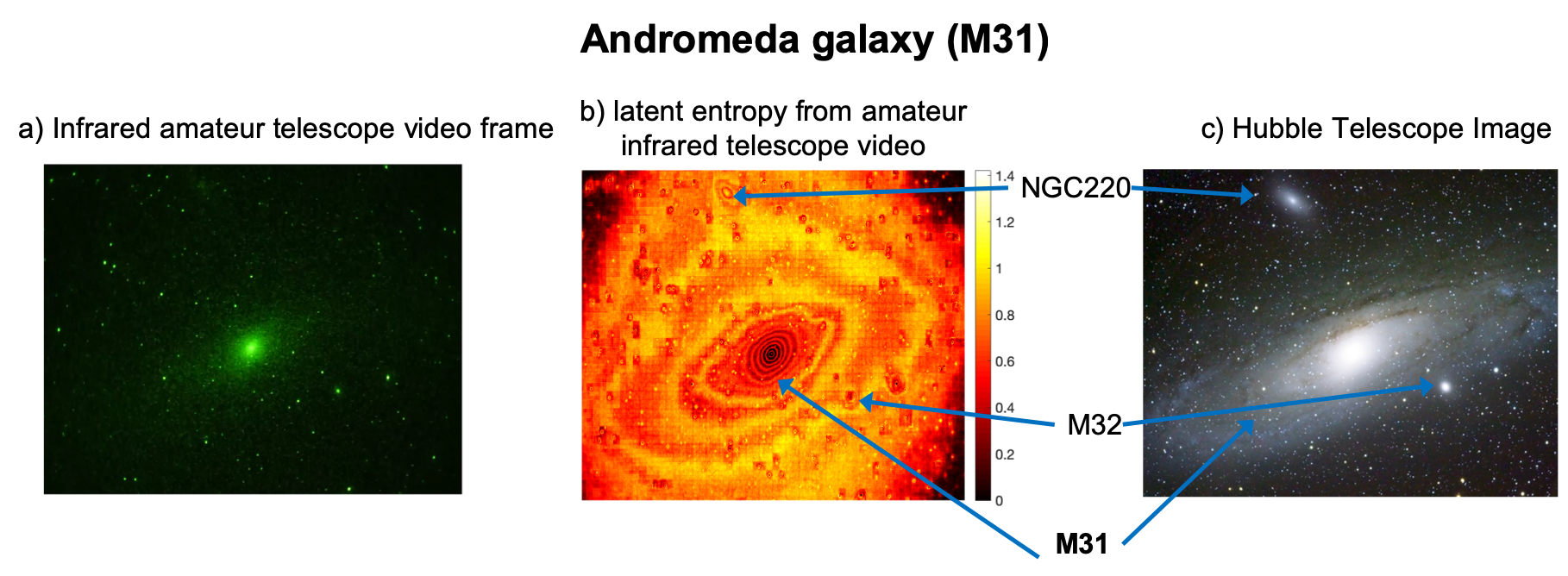}
   \caption{ Analysis of infrared video of the Andromeda galaxy (M31). The first panels shows a frame of an amateur infrared video. The second panel shows the results of latent entropy for this video. The third panel is a photo obtained by the Hubble Telescope. We notice that one can identify features observed by the Hubble telescope in the latent entropy measure.
   }
    \label{fig:andromedagalaxy}
\end{figure*} 

\section{Revealing hidden features from the video of the Andromeda galaxy}
\label{app:andromeda}
As explained in the main text, the methods introduced in this work can be used to analyse any movie data by examining the time evolution of the pixel values. To emphasize this point, we show here explicitly the results for an amateur movie of the Andromeda galaxy. 
One of the major obstacles for the earth-bound visual and near-infrared light astronomical observations is in filtering the observational data from noise. The influence of fluctuations from the atmosphere, for example, hinders the resolution of images  hence the requirement of expensive orbital instruments and complex data analysis techniques. 

In Fig.~\ref{fig:andromedagalaxy} we compare our results from analysing an amateur infrared video of the Andromeda galaxy, also identified as Messier 31, M31 or NGC 224, to observations from the Hubble Space Telescope. 
We examine a 30 second long amateur recording with 1200 color frames in the infrared part of the spectra.
An example video frame is shown Fig.~\ref{fig:andromedagalaxy}a. The movie was taken with a P43 phosphor night vision unit and a Litton 108mm/f1.5 NV lens telescope, recorded with a Panasonic GH3 camera and compressed to a lossy MP4 video storage format. From the data, the average latent entropy (middle panel) allows for resolving the structures of galaxy arms for the M31.
Comparison of this Figure to the results of the MP4-compression of the Gaussian noise data (see Fig.~\ref{fig:videocompression}) shows that the rectangular grid pattern overlaying the latent entropy results (see Fig.~\ref{fig:andromedagalaxy}c) is in fact induced by the lossy MP4 video compression - and is invisible to the common data- and video-processing tools.

\section{Latent relation measures for discretized data}
\label{app:theorydiscretel}

In the following, we provide the detailed mathematical framework for the proposed latent measures introduced in Sec.~\ref{subsec:Setup}
and prove the properties stated in Sec.~\ref{subsec:Properties}.

\subsection{Calculation of $S_K$}
\label{app:SK}
To compute the latent entropy $S_K$ for a given $K$ (Step 1) of the introduced algorithm, we define the following matrices for the transitions probability including the latent processes, $(\lambda_{K})_{ik}=\bP\left[Y(t)=y_i|L_{K}(t)=\textit{l}_k\right]$ and $(\gamma_K)_{kj}=\bP\left[L_{K}(t)=\textit{l}_k| X(t)=x_j\right]$ such that the transition matrix of Eq.~\eqref{eq:LambdaK} can be rewritten as $\Lambda_K = \lambda_K \gamma_K$.
The matrices $\gamma_K$ and $\lambda_K$ are of dimensions $n\times K$ and $K \times n$ respectively. For each value of $K$ they can be obtained by solving the negative average log-likelihood minimization problem. 
This means we search for $\lambda_K$ and $\gamma_K$ that minimize
\begin{equation}
\label{eq:LogL_full_3}
{S}_K=-\sum_{i=1}^m\sum_{j=1}^nC_{ij}\log\left({\lambda}_K{\gamma}_K\right)_{ij} 
\end{equation}
subject to the constraints 
\begin{align}
\label{eq:lambda_con1}
&({\lambda}_K)_{ik}\geq 0, \quad \sum_{i=1}^m({\lambda}_K)_{ik}= 1, \forall i,k,\\
\label{eq:gamma_con1}
&({\gamma}_K)_{kj}\geq 0, \quad \sum_{k=1}^K({\gamma}_K)_{kj}= 1, \forall k, j.
\end{align}
In Eq.~\eqref{eq:LogL_full_3}  $C_{ij}=\frac{1}{N}\sum_{t=1}^N\chi(Y(t)=y_i)\chi(X(t)=x_j)$ is the average contingency table of the data $X$ and $Y$, with $\chi$ being an indicator function. Note that $S_K \geq 0$, and as such is bounded from below.

The minimization under the constraints described above is performed in an iterative optimisation process:
First, we apply the Jensen's inequality to Eq.~\eqref{eq:LogL_full_3} to obtain an upper-bound $\hat{S}_K$ for $S_K$ with 
 \begin{equation}
 {S}_K\leq\hat{S}_K=-\sum_{i=1}^m\sum_{j=1}^n\sum_{k=1}^KC_{ij}({\gamma}_K)_{kj}\log\left(({\lambda}_K)_{ik}\right),
\label{eq:SKhat}
 \end{equation}
which can be minimised using the
{\bf Direct Bayesian Model Reduction (DBMR) Algorithm}:
\begin{itemize}
\item \underline{Initialization:} Set $\hat{S}^{(0)}_K=0$ and choose random ${\lambda}_K^{(1)}$. Set
\begin{equation}
({\gamma}_K)_{kj}^{(1)}=
\begin{cases}
1 & \mathrm{if}\, k= k^{*, (1)}\\
0 & \mathrm{else}
\end{cases}\notag
\end{equation}
for all $j$ and $k$, where 
\begin{equation}
k^{*, (1)}=\underset{k'}{\operatorname{argmax}} \sum_{i=1}^m C_{ij}\log(({\lambda}_K)^{(1)}_{ik'}).\notag
\end{equation}
Compute $\hat{S}^{(1)}_K= \hat{S}_K({\gamma}_K^{(1)},{\lambda}_K^{(1)})$ using Eq.~\eqref{eq:SKhat}
\item \underline{Iteration step:}
Iterate the following steps for $I>1$,
 \begin{enumerate}
 \item Set $({\lambda}_K)^{(I)}_{ik}=\frac{\sum_{j=1}^n C_{ij}({\gamma}_K)_{kj}^{(I-1)}}{\sum_{i'=1}^m\sum_{j=1}^n C_{i'j}({\gamma}_K)_{kj}^{(I-1)}}$ $\forall$ $i, k$
 \item Set 
 \begin{equation}
({\gamma}_K)^{(I)}_{kj}=
\begin{cases}
1 & \mathrm{if}\, k=k^{*,(I)}\\
0 & \mathrm{else}
\end{cases}\notag
\end{equation}
 for all $j, k$, where 
 \begin{equation}
k^{*,(I)} =\underset{k'}{\operatorname{argmax}}\sum_{i=1}^m C_{ij}\log(({\lambda}_K)^{(I)}_{ik'}) \notag
 \end{equation}
 \item Compute $\hat{S}^{(I)}_K= \hat{S}_K({\gamma}_K^{(I)},{\lambda}_K^{(I)})$ using Eq.~\eqref{eq:SKhat}
 \item Verify if $\|\hat{S}_K({\gamma}_K^{(I)},{\lambda}_K^{(I)})-\hat{S}_K({\gamma}_K^{(I-1)},{\lambda}_K^{(I-1)})\|$ is less than the tolerance threshold. If not, proceed to calculating $I+1$, otherwise, $\hat{S}_K^{I}$ is the desired limit to $S_{K}$.
\end{enumerate}
\end{itemize}
Switching between the optimisations for fixed iterated parameter values $\{{\gamma}_K\}$ and $\{{\lambda}_K\}$, respectively leads to a minimization of the original problem (\eqref{eq:LogL_full_3},\eqref{eq:lambda_con1},\eqref{eq:gamma_con1}), as summarized in 
Lemma 1. Note that in the main text it is $n=m$
\\
\\
\underline{\textbf{Lemma 1:}} (properties and cost of the approximate computation for $S_K$): \emph{Given two sets of categorical data  $\left\{X(1), \right.$ $\left.X(2),\dots,X(N)\right\}$ and $\left\{Y(1), Y(2),\dots,Y(N)\right\}$ 
(where for any $t$, $X(t)\in\{x_1,x_2,\dots,x_n\}$ and $Y(t)\in\{y_1,y_2,\dots,y_m\}$),
 the approximate estimates for $\{{\lambda}_K\}$ and $\{{\gamma}_K\}$ in the reduced model (\ref{eq:mast_reduced}) for a given latent dimension $K$ can be obtained via a minimisation of the upper bound $\hat{S}_K$ of the function ${S}_K$ subject to the constraints (\ref{eq:lambda_con1},\ref{eq:gamma_con1}). Solutions of this problem exist and are characterised by the discrete/deterministic optimal matrices $\{{\gamma}_K\}$ that have only elements zero and one.
 Solutions of  (\ref{eq:LogL_full_3},\ref{eq:lambda_con1},\ref{eq:gamma_con1})  can be found in a linear time, by means of the monotonically-convergent DBMR-Algorithm with a computational complexity of a single iteration scaling as 
 $\mathcal{O}\left(Kmn\right)$ and requiring no more then $\mathcal{O}\left(K(m-1)+n+mn\right)$ of memory if $N>mn$.}
 \\
\textbf{Proof:}
\begin{itemize}
\item[(a)] \underline{Existence of a solution:} 
Since ${S}_K\leq\hat{S}_K$ and ${S}_K$ (being the negative average log-likelihood function) is bounded with zero from below, function $\hat{S}_K$ is also bounded with zero from below.
Existence of a solution for the respective optimisation problem then follows straightforwardly from the boundedness of the function (\ref{eq:LogL_full_3}) and boundedness of a convex $[0,1]$--simplex domain defined by the linear constraints (\ref{eq:lambda_con1},\ref{eq:gamma_con1}) \cite{nocedal06}. Please note that this solution might not be unique.  
\item[(b)] \underline{Uniqueness of the analytical solution} wrt.\ $\{{\lambda}_K\}$ for a fixed parameter $\{{\gamma}_K\}$.
For any fixed  $\{{\gamma}_K\}$ that satisfies (\ref{eq:gamma_con1}), the problem (\ref{eq:LogL_full_3},\ref{eq:lambda_con1})
 becomes a convex minimization problem wrt.\ $\{{\lambda}_K\}$ that is subject to linear equality and inequality constraints. Deploying a standard method of Lagrange multipliers for the equality constraints only, if $\sum_{i=1}^m\sum_{j=1}^n\left({\gamma}_K\right)_{kj}C_{ij}{\neq}0$ (for all $k=1,\dots,K$) one obtains a unique optimal solution:
  \begin{eqnarray}
  \label{eq:sol_lambda}
  ({\lambda}_K)_{ik}^*&=&\frac{\sum_{j=1}^n({\gamma}_K)_{kj}C_{ij}}{\sum_{i'=1}^m\sum_{j=1}^n({\gamma}_K)_{kj}C_{i'j}},
 \end{eqnarray}
 that satisfies the inequality constraints in (\ref{eq:lambda_con1}). Therefore, it will be also a unique solution of the full problem (\ref{eq:LogL_full_3},\ref{eq:lambda_con1},\ref{eq:gamma_con1}) when $\{{\gamma}_K\}$ is fixed. Note that here and in the following we will use the "*" to indicate the optimal solution.
\item [(c)] \underline{Discrete analytical solution wrt.\ $\{{\gamma}_K\}$} for a fixed parameter $\{{\lambda}_K\}$.
For any fixed $\{{\lambda}_K\}$ that satisfies (\ref{eq:lambda_con1}), the problem (\ref{eq:LogL_full_3},\ref{eq:gamma_con1}) is a linear maximization problem (LP) with
 block-diagonal matrices of linear equality and inequality constraints. Due to this block-diagonal structure of constraints, a  solution of this  LP-problem 
  is equivalent to an independent solution of the $n$ following  LP-problems -- separately for every $j$:
 Maximize
    \begin{equation}
\label{eq:LP}
\sum_{k=1}^K\alpha_{kj}({\gamma}_K)_{kj} \\
\end{equation}
wrt.\ $({\gamma}_K)_{1j},\dots,({\gamma}_K)_{Kj}$
under the constraints
 \begin{equation}
{({\gamma}_K)_{kj}}\geq0, \quad \sum_{k=1}^K({\gamma}_K)_{kj}=1, \forall k, j.
 \end{equation}
Here $\alpha_{kj}=\sum_{i=1}^mC_{ij}\log({\lambda}_K)_{ik}$ are fixed non-positive constants when $\{{\lambda}_K\}$ is fixed. 
When $\underset{k'}{\operatorname{argmax}}{\left\{\alpha_{k'j}\right\}}$ is unique for all $j$, substituting the following expression  \begin{eqnarray}
\label{eq:gamma_ex}
({\gamma}_K)^*_{kj}&=&\left\{\begin{array}{l l}
1,&\text{if }k=\underset{k'}{\operatorname{argmax}}{\left\{\alpha_{k'j}\right\}}\\
0,&\text{else } \\
\end{array}\right.,
\end{eqnarray}
for $({\gamma}_K)_{kj}$ into Eq.~\eqref{eq:LP} provides a maximum value to the LP-functions that also satisfies the constraints, and thus  
  an optimum of the problem  (\ref{eq:LogL_full_3},\ref{eq:gamma_con1}) for fixed $\{{\lambda}_K\}$.

  When the $\underset{k'}{\operatorname{argmax}}\left\{\alpha_{k'j}
\right\}$ is not unique, i.e., when there are some $j$ for which there exists some set $k=\{k_1,k_2,\dots,k_p\}$
   such that $\alpha_{k_1j}=\dots=\alpha_{k_pj}=\underset{k'}{\max}{\left\{\alpha_{kj}\right\}}$), then the solution of (\ref{eq:LP}) is not unique. Every combination of $({\gamma}_K)_{kj}$ that 
    satisfies $({\gamma}_K)_{k_1j}+\dots+({\gamma}_K)_{k_pj}=1$, $({\gamma}_K)_{kj}{\geq}0$  $\forall$ $k$ and $j$ -- including a deterministic one where one arbitrarily-selected $({\gamma}_K)_{k'j}$ ($k'{\in}k$) is set to one and all other $({\gamma}_K)_{k''j}$ ($k''{\neq}k'$) are set to zero -  would provide an optimum of the problem  (\ref{eq:LogL_full_3},\ref{eq:gamma_con1}) for a fixed $\{{\lambda}_K\}$.
\item[(d)] \underline{Monotonic convergence of the DBMR-algorithm:}
According to step (b) and step (c) of this proof, the problem can be solved via the iterative optimisation procedure switching between the optimisations for fixed iterated parameter values $\{{\gamma}_K\}^{\left(I\right)}$ (in (c)) and $\{{\lambda}_K\}^{\left(I\right)}$ (in (b)). Iterative repetition of these two steps - starting at some arbitrarily chosen value $\{{\gamma}_K\}^{\left(1\right)}$ or $\{{\lambda}_K\}^{\left(1\right)}$ in the first algorithm iteration - will result in a monotonic decrease of the respective function value $\hat{S}_K^{\left(I\right)}$ when $I$ increases (i.e., $\hat{S}_K^{\left(I\right)}<\hat{S}_K^{\left(I+k\right)}$, where $k{\geq}1$). Since the overall problem    (\ref{eq:LogL_full_3},\ref{eq:lambda_con1},\ref{eq:gamma_con1}) is bounded from below with zero and is defined on a bounded domain, this iterations will monotonically converge to a local minimum of the function (\ref{eq:LogL_full_3}) -- dependent on the initial choice of the iteration parameters  $\{{\gamma}_K\}^{\left(1\right)}$ or $\{{\lambda}_K\}^{\left(1\right)}$.
\item [(e)] \underline{Computational iteration cost and memory scaling:}
The computational iteration complexity of the DBMR algorithm can be obtained by counting the algebraic operations during the analytical computations of the optima \eqref{eq:sol_lambda} and \eqref{eq:gamma_ex}. It  scales as $\mathcal{O}\left(K\cdot\min\{mn,N\}\right)$  in every iteration of the DBMR-algorithm. Hence, if $N>mn$ this cost becomes independent of the data statistics length $N$ and scales as $\mathcal{O}\left(Kmn\right)$. Storage of the contingency matrix $C$ - as well as of the algorithm variables $\{{\gamma}_K\}$  and  $\{{\lambda}_K\}$ requires $\mathcal{O}\left(K(m-1)+n+mn\right)$ of memory.
$\quad\quad\quad\quad\quad\quad\quad\quad\quad\quad\quad\quad\quad\quad\quad\quad\quad\square$
\end{itemize}

\subsection{Properties of the latent measures}

The following Lemma reveal some properties of the latent measures \eqref{eq:exp_SK}, which will then lead to the Theorems 1 and 2 summarizing their key properties:
\\
\\
\underline{\textbf{Lemma 2}} (monotonicity of $S_K$): \emph{Regarded as a function of $K$, the auxiliary measure $S_K$ is monotonically decreasing, i.e., $S_1\geq\dots\geq S_{K}\geq S_{K+1}\geq\dots\geq S_n$ }.
\\
\textbf{Proof:} 
Imposing additional equality constraints by setting an additional $\left(K+1\right)$-row and a $\left(K+1\right)$-column of the matrices ${\lambda}_{\left(K+1\right)}$ and ${\gamma}_{\left(K+1\right)}$ to zero, the problem (\ref{eq:LogL_full_3})-(\ref{eq:gamma_con1}) for $K$ can be written as a particular case of the problem (\ref{eq:LogL_full_3})-(\ref{eq:gamma_con1}) for $\left(K+1\right)$. Then, the same function (\ref{eq:LogL_full_3}) has to be minimized both for $K$ and for $\left(K+1\right)$. Hence, the solution $S_{K}$  of the minimization problem with more equality constraints imposed has to be less optimal then the solution of a less-constrained problem with respect to $S_{\left(K+1\right)}$. Therefore,  $S_{K}\geq S_{\left(K+1\right)}$ for any $K$ between $1$ and $n$.
$\quad\quad\quad\quad\quad\quad\square$
\\
\\
\underline{\textbf{Lemma 3}} (relation between ${S}_K$ and the latent entropy):\emph{ If  $N \rightarrow \infty$, the auxiliary function $S_K$ converges almost surely (in the sense of probability) to the differential entropy of the model (\ref{eq:mast_reduced}). }
\\
\textbf{Proof:} This can be shown by combining the law of the large numbers and the fact that the Kullback-Leibler-divergence between the distribution of the true matrix $\Lambda_K$ and the one of the parameter estimates will converge to zero almost surely. We refer to \cite{Ding06} for further details.
$\quad\quad\quad\quad\quad\quad\square$
\\
\\
Moreover from
\\
\underline{\textbf{Definition 1}} (deterministic relation between $X$ and $Y$): \emph{The Relation between the categorical variables $X$ and $Y$ is called deterministic if for every category $x_j$ of $X$ (for every $j=1,\dots,n$) there exists an $i$ ($i\in\left\{1,\dots,n\right\}$) such that $({\Lambda}_K)_{i,j}=\bP\left[Y=y_i|X=x_j\right]=1$}.
\\
\\ 
We have that
\\
\underline{\textbf{Lemma 4}} (bounds of $\bar{S}$ and $\bar{K}$): \emph{for a given categorical data $X$ and $Y$, \emph{$\bar{S}\in\left[S_n,S_1\right]$} and \emph{$\bar{K}\in\left[1,n\right]$}.}
\\
\textbf{Proof:} Since the Akaike weights $p_K$ from Eq.~\eqref{eq:AIC} are non-negative and sum-up to one, the right-hand sides of Eqs.~\eqref{eq:exp_SK} represent the convex linear combinations of $S_K$ and $K$, respectively. Then, from the monotonicity of $S_K$ and $K$ (Lemma 1) it follows that  $\bar{S}\in\left[S_n,S_1\right]$ and $\bar{K}\in\left[1,n\right]$. $\quad\quad\quad\quad\quad\quad\quad\quad\square$
\\
\\
\underline{\textbf{Lemma 5}} ($\bar{S}$ in a deterministic relation case): \emph{$\bar{S}=0$ if and only if the relationship between $X$ and $Y$ is deterministic}.
\\
\textbf{Proof:} From Lemma 2 and 4 it follows that $\bar{S}=0$ if and only if $S_n=0$, $p_n=1$ and $p_K=0$ for $K<n$. This means that  
$\bar{S}=S_n=0$ and can be only achieved if $\log\left({\lambda}_K{\gamma}_K\right)_{ij}=0$ for all $i,j$ such that $C_{ij}>0$. 
Hence,
 $({\Lambda}_K)_{ij}=1$ for all $i,j$ such that $C_{ij}>0$ and $({\Lambda}_K)_{ij}=0$ for all $i,j$ such that $C_{ij}=0$ (which corresponds to the Definition 1).$\quad\quad\quad\quad\quad\quad\quad\quad\quad\quad\quad\quad\quad\quad\quad\quad\quad\quad\quad\quad\quad\quad\square$
 \\
\\
\underline{\textbf{Lemma 6}} ($\bar{S}$ in the independent case): \emph{$\bar{S}=S_1$ and $\bar{K}=1$ if and only if $Y$ is independent of $X$}.
\\
\textbf{Proof:} From Lemma 2 and 4 it follows that $\bar{S}=S_1$ if and only if $p_1=1$ and $p_K=0$ for $K>1$. This means that the expected latent dimension $\bar{K}=1$, $({\gamma}_K)_{ii}=\left(1,1,\dots,1\right)$,  $\Lambda_K \, \Pi_X(t)= {\lambda}_1$ and Eq.~\eqref{eq:mast_reduced} takes the form 
\begin{equation}
\label{eq:mast_independent}
\Pi_Y(t)={\lambda}_1,
\end{equation}
where $\Pi_Y(t)$ is independent of $\Pi_X(t)$. Here $(\Pi_X(t))_{i} = \bP[X(t) = x_{j}] $ and $(\Pi_Y(t))_{i} = \bP[Y(t) = y_{i}] $. This implies that if $Y$ is independent of $X$, all the categories of $X$ are mapped to a single latent category, (see Fig.~\ref{fig:algorithm}).  In such a case the latent dimension is $K=1$ and the model is a Bernoulli model.$\quad\quad\quad\quad\quad\quad\quad\quad\quad\quad\quad\quad\quad\quad\quad\quad\quad\quad\quad\quad\quad\square$
\\
\\
These properties of $\bar{S}$ and $\bar{K}$ motivate the introduction of the normalized latent relation measures, through a rescaling of the measures \eqref{eq:exp_SK} to the interval $\left[0,1\right]$ as in Eq.~\eqref{eq:exp_rel} of the main text.
For these measures, we have the following theorem:
\\
\\
\underline{\textbf{Theorem 1}} (properties of relative latent entropy measure $\bar{S}_{rel}$ and relative latent dimension measure $\bar{K}_{rel}$): 
\emph{For given categorical data $X$ and $Y$, $\bar{S}_{rel}\in\left[0,1\right]$ and $\bar{K}_{rel}\in\left[0,1\right]$.  $\bar{S}_{rel}=0$ if and only if there is a deterministic relationship between $X$ and $Y$ in the sense of the Definition 1. $\bar{S}_{rel}=1$ and $\bar{K}_{rel}=0$ if and only if $Y$ is independent of $X$.}
\\
\textbf{Proof:} Statements of the Theorem 1 follow straightforwardly from Lemmas 4, 5 and 6. $\quad\quad\quad\quad\quad\quad\quad\quad\quad\quad\square$
\\
\\

Next, we consider numerical algorithms for the computation of these latent relation measures. The structure of the problem \eqref{eq:LogL_full_3}-\eqref{eq:gamma_con1} motivates the deployment of the iterative methods (e.g., of the sequential quadratic programming procedures \cite{nocedal06}), since the parameters $\{{\lambda}_K\}$ and $\{{\gamma}_K\}$ naturally separate the problem into two concave maximisation
 problems with linear equality and inequality constraints. However, following the standard procedure for this particular problem (i.e., substitution of the linear equality constraints into Eq.~\eqref{eq:LogL_full_3}, --  followed by taking the partial derivatives of the resulting function with respect to the arguments $\{{\lambda}_K\}$ and $\{{\gamma}_K\}$ and setting the obtained derivatives to zero) -- results in the nonlinear system of equations that can not be solved analytically. Moreover, the resulting system of equations does not include the inequality constraints, providing no guarantee that the obtained solutions will be non-negative. And the full numerical solution of the problem (\ref{eq:LogL_full_3},\ref{eq:lambda_con1},\ref{eq:gamma_con1}) by means of gradient-based optimisation methods would require $\left(\mathcal{O}\left(\left(2K-1\right)^3\left(2n\right)^3\right)+\mathcal{O}\left(N\right)\right)$ of operations. This means that the numerical cost of such a reduced model identification procedure will scale polynomially with the maximal possible latent dimension $n$
 \footnote{Please note that in the discrete model (\ref{eq:mast_reduced}) $K\leq n$ and the maximal possible latent dimension is $K=n$. For GMMs the maximal possible discrete latent dimension is given by the maximal number of Gaussian distributions in the mixture and is also denoted with $n$ in (\ref{eq:GMM_density})} - prohibiting an application of this method to realistic problems with large $n$. 
Another important Theorem for the measures of Eqs.~\eqref{eq:exp_SK} is
\\
\\
\underline{\textbf{Theorem 2}} (scaling of iteration cost and memory requirements in computations of latent relation measures):\emph{ For given sets of discrete/categorical data  $\left\{X(1), \right.$ $\left.X(2),\dots,X(N)\right\}$ and $\left\{Y(1), Y(2),\dots,Y(N)\right\}$ 
(where for any $t$, $X(t)\in\{x_1,x_2,\dots,x_n\}$ and $Y(t)\in\{y_1,y_2,\dots,y_m\}$) with $N>mn$, the computation of the latent relation measures $\bar{S}$, $\bar{K}$, $\bar{S}_{rel}$ and $\bar{K}_{rel}$ defined in Eqs.~\eqref{eq:exp_SK} and \eqref{eq:exp_rel} can be performed with the iteration cost that scales as $\mathcal{O}\left(mn^3\right)$. This computation would require no more than $\mathcal{O}\left((m+n)n\right)$ of memory.} 
\\
\textbf{Proof:} 
As follows from the Lemma 7,  the leading order computational  complexity for approximating $S_K$ by $\hat{S}_K$ in one DBRM iteration is  $\mathcal{O}\left(Kmn\right)$ and the iteration complexity in computing a sequence $S_1,S_2,\dots,S_n$ is:
\begin{equation}
  \begin{split}
\mathcal{O}(1mn)+\mathcal{O}(2mn)+\dots+\mathcal{O}(nmn)\\
=\mathcal{O}\left(\frac{n^2(n-1)}{2}m\right) 
=\mathcal{O}(n^3m).
\label{eq:DBRM_S}
  \end{split}
\end{equation}
In the case of a sequential  code these computations for $S_1,\dots,S_n$ would require the memory for storage of the one contingency matrix $C$ (i.e.\ $\mathcal{O}(mn)$ of memory if $N>mn$), and the DBRM variables $\{{\lambda}_K\}$ and $\{{\gamma}_K\}$  (i.e.\ up to $\mathcal{O}\left((K-1)n+K(m-1)\right)$ of memory). To leading order, the whole algorithm requires no more than  $\mathcal{O}\left((m+n)n\right)$ (since $K\leq n$). $\quad\quad\quad\quad\quad\quad\quad\quad\quad\quad\quad\quad\quad\quad\quad\quad\quad\square$
\\
\\
\underline{\textbf{Corollary 2.1}} (cost and memory requirements for latent relation measures in the case of time series analysis):\emph{ In the case of time series analysis when $Y(t)\equiv X(t+1)$ (for all $t=1,\dots,N-1$), $m=n$ and $N>n^2$, the iteration cost of computing the latent measures  $\bar{S}$, $\bar{K}$, $\bar{S}_{rel}$ and $\bar{K}_{rel}$ defined in Eqs.~\eqref{eq:exp_SK} and \eqref{eq:exp_rel} will be independent of the statistics size $N$ and the observable data dimension $D$. It will only depend on the maximal discrete latent dimension $n$ and scale as $\mathcal{O}\left(n^4\right)$, requiring  no more then $\mathcal{O}\left(n^2\right)$ of memory.  } 
\\
\textbf{Proof:} Statement of the Corollary follows from the Lemma 1 and the Theorem 2 when $m=n$ and $N>n^2$. $\quad\quad\quad\quad\quad\square$

In the following we show that computing the expected latent entropies from the GMM measures $S_1,\dots,S_n$ defined in (\ref{eq:GMM_S_n})  (with $n$ being the maximal allowed discrete latent dimension)  will grow linearly with the data dimension $D$ and the statistics size $N$ - and would have an iteration cost scaling of $\mathcal{O}\left(n^2ND\right)$, requiring $\mathcal{O}\left(n(N+D)\right)$ of memory.

\section{Comparison to other statistical measures}
\label{app:comparison}

In this part of the appendix we compare the introduce measures to other latent machine learning tools, in particular to GMMs, which with almost 1 Mio. citations (according to the Google Scholar) belong to the most popular latent inference methods \cite{schnatter06,Greggio2012}. For a given data sequence $X=\{X(t=1),X(t=2),\dots,X(t=N)\}$ (where $X(t)$ is a $D$-dimensional Euclidean vector for every $t$), GMMs fit a mixture of $n$ (multivariate, i.e., $D$-dimensional) Gaussian distributions  
\begin{equation}
\label{eq:GMM_density}
\bP_{GMM}\left[X(t)\right]=\sum_{i=1}^np_i\,\mathcal{N}\left(X(t);\mu_i,\Sigma_i\right),
\end{equation}
where $p_i$ is a relative weight of the Gaussian $i$ (with the mean vector $\mu_i$ and covariance $\Sigma_i$) in the mixture, such that $p_i\geq 0$ (for all $i$ ) and $\sum_ip_i=1$. Hereby, for every data point $X(t)$ one assumes the presence of the categorical latent variable $L_t$, taking values from a finite set of $n$ values $1,\dots,n$, indicating which of the Gaussian distributions from the mixture (\ref{eq:GMM_density}) is actually responsible for the generation of this particular $X(t)$.

Different variants of the Expectation Maximisation algorithm (EM) have been developed to find the optimal GMM parameters $\left(p_i,\mu_i,\Sigma_i\right), i=1,\dots,n$ for a given data $X=\{X(t=1),X(t=2),\dots,X(t=N)\}$ and with a fixed number of mixture components $n$ \cite{schnatter06,Greggio2012,pinto15}. EM also provides the estimates of probabilities $({\gamma}_n)_{it}=\bP\left[L_t=i|X(t)\right]$ for a latent process $L_t$ to be in the latent state $i$ at the instance $t$.  In image processing applications one frequently uses the negative average log-likelihood $S_n$ of the fitted model  (\ref{eq:GMM_density}) as a feature intensity measure \cite{zoran11,Greggio2012,bouman18,bouman19}:\\
\begin{equation}
\begin{split}
S_n&=-\frac{1}{N}\log\bP_{GMM}\left[X\right]=-\frac{1}{N}\log\prod_{t=1}^{N}\bP_{GMM}\left[X(t)\right]\\
&=-\frac{1}{N}\sum_{i=1}^n\sum_{t=1}^{N}({\gamma}_n)_{it}\left[\log(p_i)+\log\left(\mathcal{N}\left(X(t);\mu_i,\Sigma_i\right)\right)\right].
\end{split}
\label{eq:GMM_S_n}
\end{equation}
To reduce the computational cost of the EM algorithm, one frequently restricts the covariance matrices $\Sigma_i$ 
to be diagonal  \cite{zoran11,bouman18,bouman19}. The following Lemma summarises the properties of cost and memory scalings for the computations of $S_1,S_2,\dots,S_n$  with basic variants of GMMs used in the image processing.
 \\
 \\
\underline{\textbf{Lemma 7}} (GMMs scaling for latent relations computation): \emph{for the given observational data $X=\{X(t=1),X(t=2),\dots,X(t=N)\}$ of dimension $D$, the iteration complexity for the sequence of negative log-likelihoods computations  $S_1,\dots,S_n$ defined in Eq.~\eqref{eq:GMM_S_n} with diagonal Gaussians scales in the leading order as $\mathcal{O}\left(n^2ND\right)$ and requires $\mathcal{O}\left(n(N+D)\right)$ of memory}.
\\
\textbf{Proof:} 
An expectation step of the EM algorithm provides updates of the latent process probabilities ${\gamma}_n$.  For a fixed $K$ the cost of this operation is $\mathcal{O}(KND(D-1)/2)$ for non-diagonal Gaussians and  $\mathcal{O}(KND)$ for the diagonal ones. Computational cost of the single maximisation step (where the mixture model parameters $\left(p_i,\mu_i,\Sigma_i\right), i=1,\dots,n$ are updated), scales as  
$\mathcal{O}\left((K+2KD)N\right)$
 in the diagonal Gaussian case. Hence, the leading order computational  complexity for $S_K$ in one EM iteration is  
 $\mathcal{O}(KND)$ 
 and the iteration complexity in computing a sequence $S_1,S_2,\dots,S_n$ is:
\begin{equation}
  \begin{split}
\label{eq:GMM_S}
\mathcal{O}(1ND)+\mathcal{O}(2ND)+\dots+\mathcal{O}(nND)\\
=\mathcal{O}\left(\frac{n(n-1)}{2}ND\right)=\mathcal{O}(n^2ND).
\end{split}
\end{equation}
In the case of a sequential  code these computations for $S_1,\dots,S_n$ would require the memory for storage of the data $X$ ($\mathcal{O}(ND)$ of memory), the latent probailities  ${\gamma}$ (up to $\mathcal{O}(nN)$ of memory) and the GMM parameters  $\left(p_i,\mu_i,\Sigma_i\right), i=1,\dots,n$ (up to $\mathcal{O}\left(n(1+2D)\right)$ of memory). When $N$ is larger than $n$ and $D$,  this results in the leading order memory scaling of $\mathcal{O}\left(n(N+D)\right)$.  $\quad\quad\quad\quad\quad\quad\quad\quad\quad\quad\quad\quad\quad\quad\quad\quad\quad\square$

\end{document}